\documentclass[a4paper,12pt]{article}
\usepackage{citesort}
\usepackage{ae}
\usepackage[square]{natbib}
\usepackage{graphicx}
\usepackage{amssymb}

\title{Theoretical study of the influence of confinement and channel
  blocking on adsorption and diffusion of n-butane in silicalite-1}
\author{Barbara Jagoda-Cwiklik, Lukasz Cwiklik, Marek Frankowicz
\\
\\Department of Theoretical Chemistry, Faculty of Chemistry
\\Jagiellonian University
\\Krakow, Poland
}

\begin{document}

\maketitle

\begin{abstract}
  Dynamic Monte Carlo simulations for the open coarse-grained model of
  MFI type zeolite were used to study the dynamics of adsorption and
  diffusion of n-butane in silicalite-1. We demonstrated the influence
  of the confinement of the structure of zeolite channels on both the
  dynamics of the adsorption process and the maximum loading of
  adsorbate. We showed that the confinement and channel blocking limit
  the adsorption and desorption processes. Moreover, they cause the
  maximum loading in the zeolite structure for moderate pressures to
  be higher than the one predicted by the Langmuir model for a flat
  and homogeneous system.
\end{abstract}


\section{Introduction} \label{Intro}

Zeolites are microporous materials used in many chemical processes \cite{Sherman, Ghobarkar}. Due to unique physical and chemical properties zeolites are used as catalysts in processes such as catalytic cracking and dewaxing \cite{Weitkamp1, Weitkamp2, Kunga, Corma}. Other important applications of zeolites are separation and purification processes, where zeolites are employed as molecular sieves \cite{Ackley, Chau, Piera}. Diffusion and adsorption are phenomena crucial to understand processes occurring in zeolite materials and their understanding is very desirable for practical applications of zeolites.

Diffusion of adsorbed particles in zeolites is currently intensively investigated both experimentally \cite{Karger2003, Ruthven, Jobic, Song, Onyestyak} and theoretically \cite{Auerbach2000, vanSanten, Krishna2002, Takaba, Szabelski, Leroy}. From the macroscopic point of view, the diffusion can be either induced by concentration gradients of an adsorbate (transport diffusion) or can proceed in the absence of such gradients (self-diffusion). Transport diffusion is a non-equilibrium process whereas self-diffusion takes place under equilibrium conditions. Because of confinement and microporosity of zeolites, i.e. the fact that the sufficiently large particles diffusing in zeolite pores cannot pass each other, the requirements of Fickian diffusion are not fulfilled during the diffusion in zeolite channels and such diffusion often does not obey Fick's first law or obeys it only approximately (such a process is called anomalous diffusion) \cite{Karger2003, Karger_book}. Additionally, the complicated topology of the pore structure in some zeolites, i.e. the presence of interconnections between channels, influences the diffusion process and makes the investigation of diffusion in such materials highly difficult.

Adsorption in zeolites is always related to diffusion. At the beginning of a process, a zeolite material is placed in a system containing gas. Gas particles impact the zeolite crystallite and some of them are adsorbed at the openings of the zeolite channels. Then adsorbed particles can diffuse into the pore structure. The inverse process is also possible. An adsorbed particle diffusing in the pore structure can migrate to a channel opening and desorb to the gas phase. The whole process can reach equilibrium or a steady-state but the diffusion in pores still proceeds. We must emphasize that transport diffusion is important, especially during the adsorption process. Before reaching equilibrium or steady-state transport diffusion is one of the processes governing the behavior of the system and it can have an impact on the equilibrium behavior. The confinement of the zeolite structure causes the diffusion to influence the adsorption process, even if diffusivity in the pores is quick in comparison with other processes. In the present paper we will show examples of this influence.

Three main groups of methods are used in theoretical computational studies of diffusion and adsorption in zeolites. Quantum chemical computations, including DFT methods, provide one with a detailed description of energetic aspects of elementary processes. Using these methods one can obtain heights of diffusion or adsorption barriers for a given adsorbate particle in certain adsorption centers in a zeolite channel, for example \cite{vanBekkum_book, Dabrowski}. However, in practice, the usage of such methods is limited to investigations of interactions between a small fragment of the zeolite structure and a single particle of the adsorbate.

Molecular Dynamics (MD) simulations are the second group of computational methods. They give the detailed dynamics of processes proceeding in the channels. But the computational effort necessary to carry out such simulations makes it impossible to investigate processes with characteristic time scales in the range of the hundreds of nanoseconds needed, for example, to describe the diffusion of benzene in silicalite \cite{Takaba, vanBekkum_book, Nascimento, Krishna2000, Krishna2003}.

Monte Carlo simulations are the third group of these methods. Most often authors use Configurational Biased Monte Carlo (CBMC) to examine adsorption in porous materials \cite{vanBekkum_book, Krishna2001, Krishna2003, Krishna2001a}. A few years ago, a new variant of the Monte Carlo method - Dynamic Monte Carlo (DMC) was introduced \cite{Fichthorn1991, Jansen1999}, and recently has been widely used, especially to study diffusion in zeolites (see examples in \cite{Takaba, Krishna2003}). CBMC enables one to investigate systems only in thermodynamic equilibrium, whereas DMC methods are suitable for investigating the dynamics of processes, both in equilibrium and non-equilibrium conditions.

As mentioned earlier, transport diffusion, which is a non-equilibrium phenomenon, is particularly important to understand adsorption in confined systems. Therefore DMC simulations are well suited to study such problems. In this paper we investigate the diffusion and adsorption of alkanes in zeolites using DMC simulations and taking n-butane and silicalite-1 as an example. This topic has been previously studied with CBMC simulations and DMC methods \cite{Takaba, Krishna2001}. However, in the present work, we propose a simulation scheme which allows one to consider directly transport diffusion simultaneously with self-diffusion and adsorption. Using this model we study the influence of diffusion on the adsorption process. We show that this influence is caused by the confinement of the pore structure in zeolites.

\section{Description of the model} \label{ModDes}

\subsection{Simulation lattice} \label{SimLatt}

In our DMC simulations we employed a three-dimensional coarse-grained model of ZSM-5 zeolite described by Trout et al. in \cite{Trout}. This model assumes that the zeolite pore structure can be modelled as a system of intersecting channels. Therefore, two kinds of sites exist in this model: channels and intersections. The channels and intersections are connected in such a way that they reflect the actual zeolite structure: every channel connects two intersections and every intersection joins four channels. The unit cell in this model consists of 24 lattice points (16 channels and 8 intersections). We assumed, after Trout, that in the individual channel or intersection only one adsorbate particle can be placed. This assumption is reasonable taking into account the size of n-butane particles and the diameters of pores in silicalite-1. Adsorptive properties of channels and intersections are assumed to be the same. This model includes the topology of the pore structure of ZSM-5 properly and as it was shown in \cite{Krishna2002}, it also gives values of calculated self-diffusivities, which are in good agreement with experimental data. In this coarse-grained model two important assumptions were made. Particles of adsorbate were treated as points, i.e. their geometrical and structural properties were neglected. Adsorbate-adsorbate interactions were also neglected except for the fact that two particles could not occupy the same lattice site (it can be treated as a kind of a coarse-grained hard-spheres potential).

We modified Trout's model in order to study transport diffusion and adsorption. We assumed that we simulated a finite-size zeolite crystallite surrounded by the gaseous adsorbate. We also assumed that the open boundary conditions of our simulation lattice, i.e. terminations of zeolite micropores in our model, were open to the gas phase. In our model particles of the adsorbate could adsorb at the openings of zeolite pores and then diffuse into them. Particles could also migrate from inside of simulation lattice and desorb from openings of micropores to the gas phase. Therefore we were able to simulate the filling of zeolite crystallite with the adsorbate and non-equilibrium diffusion of adsorbed particles inside the system of pores.

\subsection{Processes} \label{processes}

We considered three processes taking place during our simulations: adsorption from and desorption to the gas phase and diffusion inside the pores. We used the jump-diffusion model to simulate the diffusion process. This model is commonly used in coarse-grained simulations because it assumes that adsorbate particles occupy only adsorption sites (channels or intersections in our case) and the diffusion of adsorbate particles is modelled as jumps of these particles between neighboring adsorption sites. We assumed that the diffusion was activated and that a particle needed a certain amount of energy to diffuse. Desorption of an adsorbed particle in this model was treated as hopping of a particle from the adsorption site located in a channel opening to the gas phase, i.e. the energy barrier needed to be crossed during the desorption and the pre-exponential factor of this process were assumed to be the same as in the case of diffusion because in both processes a particle needed to leave an adsorption site. In general, this assumption is not valid but one need detail microscopic barriers (for example, from quantum calculations) to distinguish between hopping and desorption. Therefore, the same barriers describing these two processes were taken in this work as a rough approximation because microscopic data are not available.

In our simulations the adsorption of particles from the gas phase at the openings of pores was modelled as a non-activated process and depended on the pressure of gaseous adsorbate only, whereas the desorption from openings to the gas phase was activated. 

\subsection{Algorithm} \label{sec:algorithm}

Simulations were performed using a three-dimensional cubic lattice. The simulation lattice was created by multiplying the unit cell which consisted of 64 cubic lattice sites. The simulation lattice had to correspond to the topology of actual zeolite pore structure, therefore in our unit cell only 24 among 64 sites were accessible for particles. As mentioned previously, among these 24 sites, 16 were assumed to be channels and 8 to be intersections and these sites were chosen in a way to give the proper topology. The unit cell in our simulation lattice corresponded to 2 unit cells of actual ZSM-5 zeolite.

The following algorithm has been used during our simulations:

\begin{enumerate}

\item \label{alg:assumptions} The pressure of the adsorbate ($p$) and the system temperature ($T$) were chosen. For each process, pre-exponential factors ($A_i$) and energy barriers ($E_i$) were assumed. Rate constants of diffusion and adsorption were calculated according to the Arrhenius' law: $k_i=A_iexp\left(\frac{E_i}{kT}\right)$, where $k$ was Boltzmann factor.

\item \label{alg:prob_calculations} 
Probabilities of processes were calculated using the values of rate constants, pressure and numbers of particles taking place in the given process ($N_i$) according to the equation: $p_i=\frac{k_i N_i}{R}$, where $R=k_{diff} N_{ads} + k_{ads} N_e$, where $k_{diff}$ -- diffusion rate constant, $N_{ads}$ -- number of particles 'adsorbed' on the lattice, $k_{ads}$ -- adsorption rate constant (equal to $p$ -- adsorbate pressure in the gas phase), $N_e$ -- number of unoccupied channel openings

\item \label{alg:process_choice}
According to probabilities a process was randomly chosen:
\begin{itemize}
\item adsorption of a particle from the gas phase
\item diffusion of an adsorbed particle 
\end{itemize}

\item \label{alg:adsorption_was_chosen}
If adsorption was chosen, a lattice site corresponding to an empty opening of a channel was chosen randomly and a particle was placed in this site

\item \label{alg:diffusion_was_chosen}
If diffusion was chosen:
\begin{itemize}
\item a lattice site occupied by a particle was chosen randomly
\item one of 6 directions for the move was chosen randomly
\item if the lattice site adjacent to the chosen one and located in the chosen direction was empty, the particle was replaced to this site (if the chosen site was a channel opening and the move was directed outside the lattice the particle was 'desorbed' to the gas phase)
\end{itemize}

\item \label{alg:time_increment}
After successive realization of adsorption or diffusion, time was increased according to:
$\Delta t=R^{-1}$

\item \label{alg:repeeat_until_sstate}
Steps \ref{alg:prob_calculations} - \ref{alg:time_increment} were repeated until a stationary state was reached

\item \label{alg:prssure_increment}
If we wanted to obtain the adsorption isotherm the pressure $p$ was increased and steps \ref{alg:prob_calculations} - \ref{alg:time_increment} were repeated until a new stationary state was reached

\end{enumerate}

The parameters needed to perform a simulation were: the size of simulation lattice, pressure of adsorbate, system temperature, pre-exponential factor and the energy barrier of hopping between two adjacent adsorption sites.  

These parameters in our simulations were chosen to correspond to experimental data for investigated systems and the choice of these parameters is described in the next section.

\subsection{Calculation of pressure} \label{sec:pressure}
In our simulation algorithm we had a parameter $p$ called the pressure. But in fact it was the rate constant of adsorption, i.e., the frequency of adsorption of molecules from the gas phase at the channels openings. In order to connect this parameter to values of actual pressure we used the Hertz-Knudsen equation for the frequency of collisions of gas particles at the surface area $A$:

$\nu=\mathbb{P} A \left(2 \pi m k T \right)^{-1/2}$

where $\mathbb{P}$ is the actual pressure measured in Pa. We also took into account the ratio ($a$) of the surface area of the openings of channels to the whole surface area of the facets of the simulated zeolite crystallite. This ratio for ZSM-5 is approximately equal to 0.1. Additionally, we assumed that only those particles which are properly oriented can adsorb at the adsorption centers in the channels openings. The proper orientation means the geometrical orientation that allows an n-butane particle to get into a zeolite channel. We assume the average diameter of a channel equal to 5,5 \AA{} and the effective length of an n-butane particle equal to 8,2 \AA{}. The integration over all orientations of n-butane particles gave us as a result that about 1\% of n-butane particles is properly oriented. That is why we introduced the ratio of properly oriented particles $b=0.01$. Finally, we get the relation between the actual pressure $\mathbb{P}$ and our simulation parameter $p$ at temperature $T$:

$\mathbb{P} = \left( a b A \right)^{-1} \left( 2 \pi mkT \right)^{1/2} p$

\section{Simulations and results} \label{sec:Results}

\subsection{Parameters of simulations} \label{sub:param_of_sim}

In order to simulate adsorption and diffusion of n-butane in silicalite-1 we took the pre-exponential factor of the hopping of adsorbate between adjacent adsorption sites to be equal to $3\cdot10^{9}$ s$^{-1}$ according to Trout et al. \cite{Trout} and, based on the results of van den Begin et al. \cite{Begin1989}, $21$ kJ/mol as the energy barrier for this process. Values of pressure and temperature were assumed in such a way that they enable comparison between our results and experimental data, i.e. pressure was in the range $10^{0} \div 10^{5}$ Pa and temperature varied from 303 to 773 K. We used simulation lattices consisting of: $16^3$, $24^3$, $32^3$ and $48^3$ sites (as mentioned in Sec. \ref{sec:algorithm}, to mimic topological properties of the system of channels in the zeolite, some of lattice sites in each simulation were not accessible to particles). These lattice sizes corresponded to 128, 432, 1024, 3456 unit cells of actual silicalite-1 and to approximate crystallite sizes (in \AA{}): 161$\times$79$\times$53, 241$\times$118$\times$79, 321$\times$158$\times$105, 482$\times$237$\times$158.

\subsection {Simulations of adsorption dynamics} \label{sub:dynamics}

\begin{figure}
\centering
\includegraphics [width=0.9\textwidth] {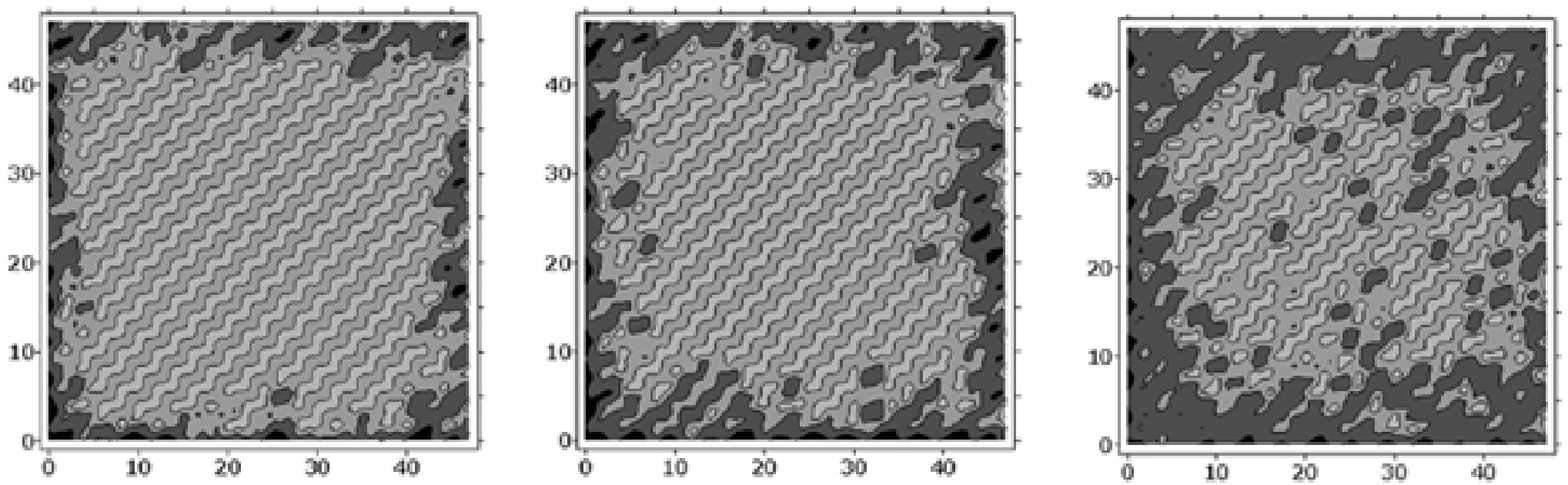}
(a)
\includegraphics [width=0.9\textwidth] {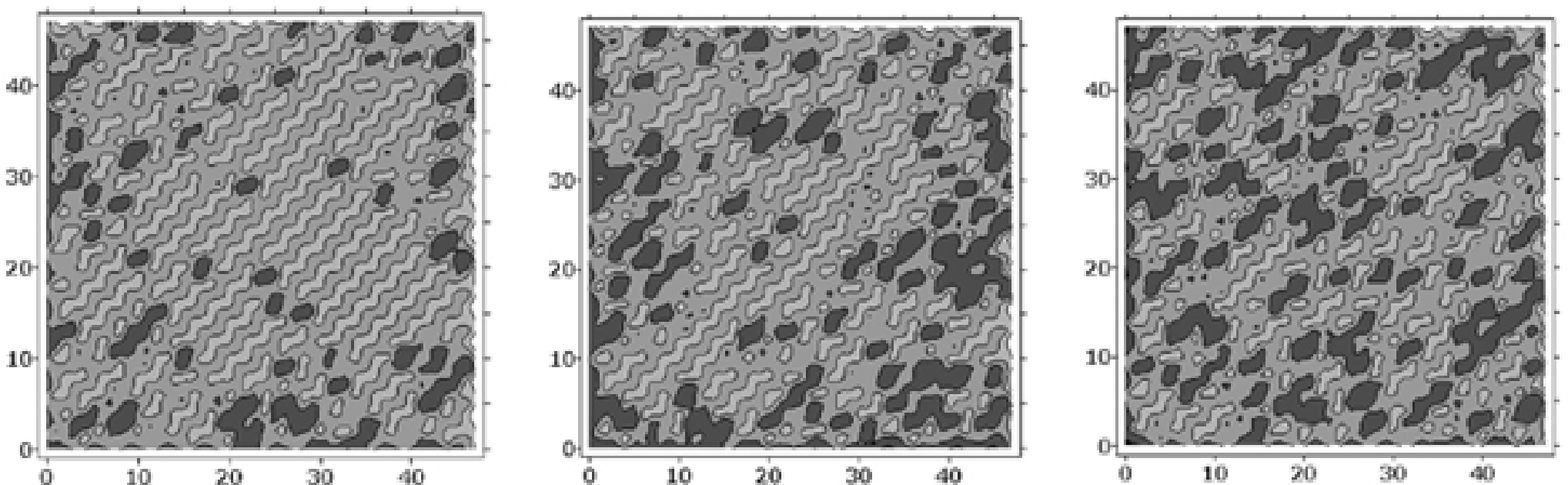}
(b)

\caption{Contour maps of cross-sections of the simulation lattice respectively for 2, 4 and 8 $\mu$s from the beginning of the simulation. Lattice size: $48^{3}$ sites, $p=10^{4}$ Pa, (a) T = 473 K, (b) T = 773 K} \label{fig:crosssections}
\end{figure}

In order to investigate the dynamics of adsorption of n-butane in silicalite-1 we performed a series of simulations, each starting from the empty lattice, and we analyzed the filling of the lattice with the adsorbate particles as a function of time for different pressures and temperatures. Fig. \ref{fig:crosssections} shows contour maps of cross-sections made in the middle of simulation lattice for different values of time counted from the beginning of the simulation for temperature 473 K (a) and 773 K (b). According to the simulation algorithm, each crystallite facet was accessible for the gaseous adsorbate, therefore adsorbate particles were able to diffuse to adsorption sites in this cross-section due to diffusion from the four edges of that cross-section and diffusion from the above and below lattice layers. Contour maps were smoothed, therefore they can be interpreted as showing the density of various centers in the simulation lattice. Light-gray areas represent empty space in the zeolite structure, semi-gray correspond to the channels and dark-gray represent channels with adsorbed particles. Using such cross-sections one can watch the filling of zeolite and analyze it, especially qualitatively.

\begin{figure}
\centering
\includegraphics[width=0.9\textwidth]{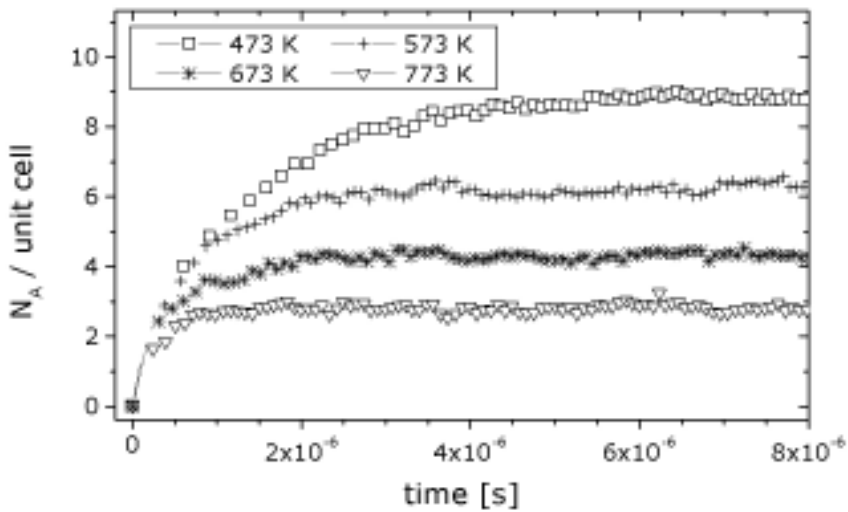}
\caption{Number of adsorbed particles per unit cell vs. time for different temperatures. Lattice size: $16^{3}$ sites, $p=10^{4}$ Pa} \label{fig:filling_vs_t_temp}
\end{figure}

Fig. \ref{fig:filling_vs_t_temp} shows the loading vs. time for different temperatures. In the initial phase ($<0.3\cdot10^{-6}$ s) the rate of filling is comparable for each temperature, but later the filling begins to slow down and this decrease in the rate of filling occurs earlier for higher temperatures. Finally, each system reaches a steady-state. The maximum loading decreases with increasing temperature. This dependence of the steady-state loading on temperature is in agreement with both experiments and a Langmuir model of adsorption.

\begin{figure}
\centering
\includegraphics [width=0.9\textwidth] {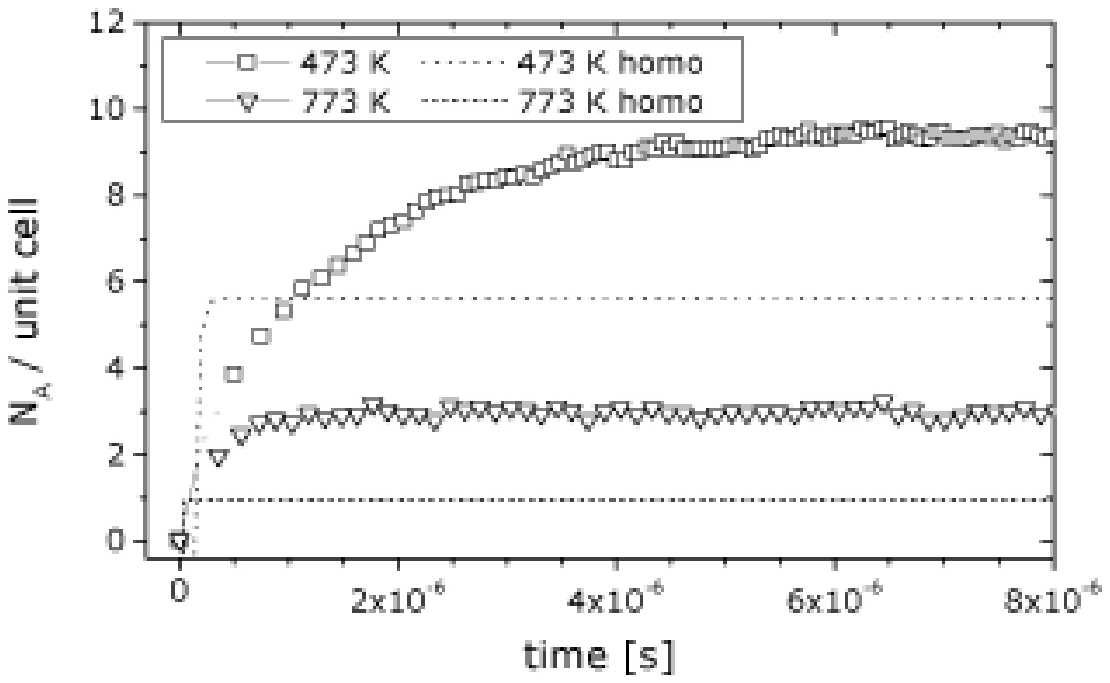}
\caption{Number of adsorbed particles per unit cell vs. time for different temperatures -- comparison of the zeolite model with the Langmuir model. Lattice size: $16^{3}$ sites, $p=10^{4}$ Pa} \label{fig:filling_vs_t_temp_hom}
\end{figure}

Fig. \ref{fig:filling_vs_t_temp_hom} shows loading vs. time for temperature 473 K and 773 K obtained from simulations and calculated using the Langmuir model of adsorption-desorption corresponding to adsorption on a flat homogeneous surface with values of rate constants and pressure equal to those taken in the simulations. The comparison between simulations and the homogeneous case shows that in the homogeneous case a maximum loading is achieved much faster than in the zeolite. The value of loading in the zeolite increases slowly, however, after a certain time (for example, at about 1.5 $\mu$s at the temperature 473 K) it becomes higher than in the homogeneous case and, finally, the steady-state loading in the zeolite surpasses the one predicted by the Langmuir model.

We can conclude that this effect is caused by the confinement of adsorbed particles in the system of zeolite channels. At the beginning of the adsorption process only those adsorption sites that are located in the openings of channels are accessible for adsorbing particles. Furthermore, an adsorbate particle has to wait until a particle adsorbed earlier at the opening either desorbs or diffuses into a channel and also the particles block one another during diffusion in the channels. Therefore, in the first phase of the adsorption process, the loading is lower than in the case of a flat homogeneous surface. During the process the particles diffuse into the channels and the loading (in comparison with the Langmuir model) increases slowly. Then a dynamic equilibrium between adsorption and desorption occurs, and, because of confinement, the desorption process is limited by the same factors as adsorption is, i.e., desorption can proceeds only at the openings of channels and the transport of adsorbate particles from the bulk of zeolite to these openings is limited because of channel blocking. This confinement effect results in a steady-state loading for adsorption in the zeolite that is higher than in the Langmuir model. The differences between maximum loading in the zeolite and homogeneous model is discussed also in the Sec. \ref{sub:isotherms}.

\begin{figure}
\centering
\includegraphics [width=0.9\textwidth] {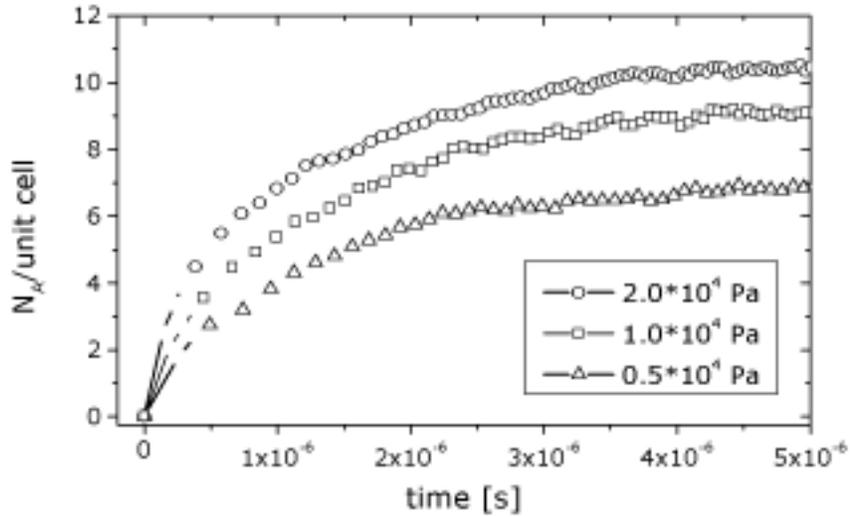}
\caption{Number of adsorbed particles per unit cell vs. time for different pressures. Lattice size: $16^{3}$ sites, T = 473 K} \label{fig:filling_vs_t_press}
\end{figure}

In Fig. \ref{fig:filling_vs_t_press} the loading vs. time for different pressures is shown. Both the rate of loading and the maximum loading increase when the pressure rises. This form of pressure dependence on the maximum loading is experimentally and theoretically obvious and it will also manifest itself in simulations of adsorption isotherms.

\subsection {Simulations of adsorption isotherms} \label{sub:isotherms}
\begin{figure}
\centering
\includegraphics [width=0.9\textwidth] {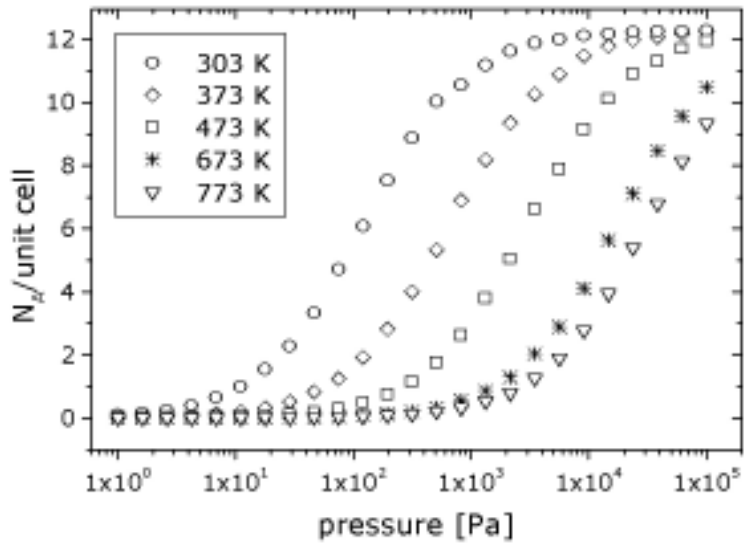}
\caption{Number of adsorbed particles per unit cell vs. pressure. Lattice size: $16^{3}$ sites} \label{fig:isotherms}
\end{figure}

As discussed earlier in Sec. \ref{Intro} the adsorption properties of zeolites strongly depend on the dynamics of transport diffusion as described in the previous section. Therefore this dynamics influences also the adsorption isotherms. Fig. \ref{fig:isotherms} presents adsorption isotherms simulated for different values of temperatures. Increasing temperature causes a shifting of isotherms toward higher pressure and this behavior is in qualitative agreement with experimental data \cite{Zhu}.

\begin{figure}
\centering
\includegraphics [width=0.9\textwidth] {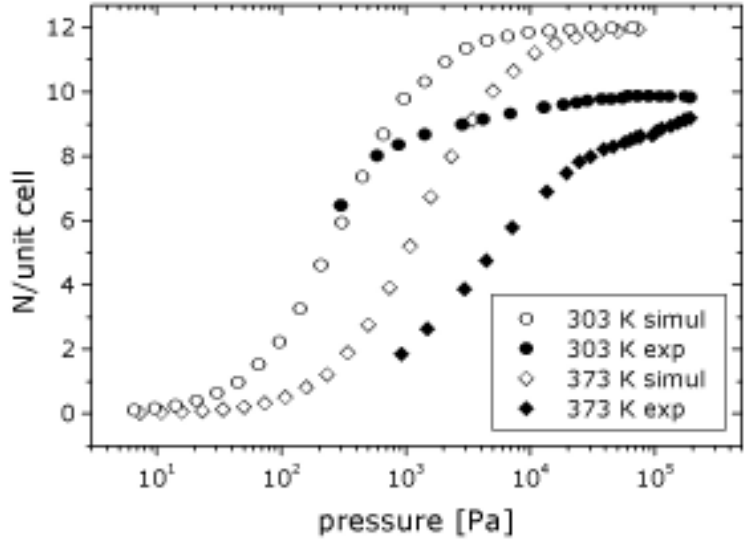}
\caption{Number of adsorbed particles per unit cell vs. pressure -- comparison with experimental data. Lattice size: $16^{3}$ sites} \label{fig:isotherms_exp}
\end{figure}

A quantitative comparison between simulated and experimental adsorption isotherms of n-pentane in silicalite-1 is shown in Fig. \ref{fig:isotherms_exp}. At both temperatures simulated isotherms differ from experimental ones in the range of high pressures. The difference is stronger for T=373K. The simulated maximum loading is higher than the loading in experiments. We must stress, that the differences can be caused by the simplicity of the zeolite model. The overestimation of maximum loadings may be caused by the assumption of the hard-spheres potential and the neglecting the long-range adsorbate-adsorbate interactions. The other source of disagreements of simulated and experimental isotherms may be our rough method of calculating the absolute value of pressure. In this method we neglect the adsorption energy barrier, for example, and we calculate a sticking coefficient only by simple geometrical considerations. Taking this into account, we can conclude, that the quantitative agreement between simulated and experimental isotherms is quite good but some corrections of the model assumptions should be done especially to improve the value of maximum loading. The stronger difference between experimental and simulated isotherms at 373K in comparison with 303K may be caused be the fact that in higher temperatures the adsorbate particles can easier bypass one another at intersections and the assumption that such bypassing is not possible may fail.

\begin{figure}
\centering
\includegraphics [width=0.9\textwidth] {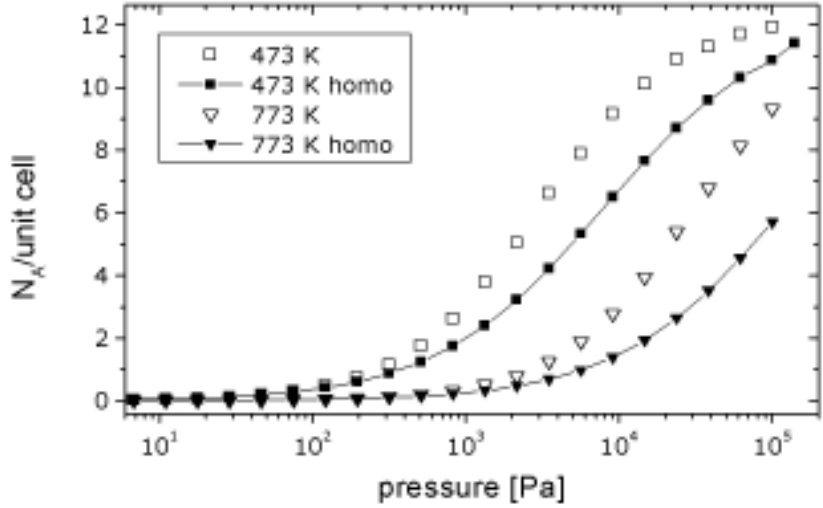}
\caption{Number of adsorbed particles per unit cell vs. pressure -- comparison of the zeolite model with the Langmuir model. Lattice size: $16^{3}$ sites} \label{fig:isotherms_hom}
\end{figure}

In order to investigate the impact of confinement on adsorption isotherms we compared simulated isotherms with isotherms calculated for the Langmuir model. This comparison is presented in Fig. \ref{fig:isotherms_hom}. For small values of pressure (below about $10^2$ Pa) the amounts of adsorbed particles calculated using the Langmuir model and simulated for the zeolite are almost equal. In the intermediate range of pressures (between $10^2$ Pa and $5\cdot10^4$ Pa for T = 473 K) the amount of particles adsorbed in the zeolite material is higher than for the homogeneous model. The reason for this effect is the confinement of adsorbate in the system of channels and the limiting conditions for the desorption processes. The dynamics of this effect have been presented in Fig. \ref{fig:filling_vs_t_temp_hom} and discussed in the previous section. In the limit of high pressures the amount of adsorbed particles simulated for the zeolite and calculates using Langmuir model will be equal.

\section{Conclusions} \label{sec:Conclusions}

The confinement of the structure of channels strongly influences both the dynamics and the steady-state properties of adsorption of n-butane in silicalite-1. The rate of adsorption in the zeolite is lower than the rate predicted by the Langmuir model for a flat and homogeneous surface under the same conditions. During adsorption, the particles of gaseous adsorbate can adsorb directly only at the openings of channels, therefore the adsorption process is limited. The second limiting factor in adsorption is channel blocking by particles adsorbed within the zeolite. Such blocking makes channels unavailable to adsorption for succeeding particles. However, these two factors limit the desorption process in the late stage of adsorption as well, therefore the final loading can be higher than in the case of a flat surface.

The comparison between simulated isotherms and ones calculated using the Langmuir model shows that the effect of increasing loading in the zeolite structure depends on the pressure. The effect manifests in the range of moderate pressures. At high and low values of pressure, the loading (or coverage) on the flat surface with the same number of adsorption sites would be the same as in the zeolite.

The simulated adsorption isotherms are in quite good agreement with experimental ones taking into account the simplicity of our open coarse-grained model used in simulations. Some corrections of the model assumptions should be done to improve the value of maximum loading. We suppose, it can be done by including adsorbate-adsorbate interactions. Our model can be further used for investigations of adsorption dynamics and confinement effects for other zeolite -- adsorbate systems. This work shows also the usefulness of Dynamic Monte Carlo method to study the processes occurring in such systems. The relatively small computational effort to study such systems in comparison with Molecular Dynamics or Configurational Biased Monte Carlo methods makes Dynamic Monte Carlo simulations a leading technique to study non-equilibrium adsorption and diffusion phenomena in heterogeneous surface systems.

\section{Acknowledgements} \label{Acknowledgements}
B. J-C. would like to thank Dr. Waclaw Makowski for fruitful discussions and helpful comments.

\end{document}